\documentclass[twocolumn,showpacs,nofootinbib,10pt,aps,pra]{revtex4-1}
\usepackage{amsmath,amssymb,graphicx,color,todonotes}

\newcommand{\braket}[2]{\langle#1|#2\rangle}

\newcommand{\bc}{\begin{cases}\begin{aligned}}
\newcommand{\ec}{\end{aligned}\end{cases}}
\newcommand{\eq}{\begin{equation}}
\newcommand{\fine}{\end{equation}}
\newcommand{\beq}{\begin{equation}}
\newcommand{\eeq}{\end{equation}}

\newcommand{\uno}{\leavevmode\hbox{\small1\normalsize\kern-.33em1}}

\newcommand{\dd}{{\rm d}}

\newcommand{\casi}{\begin{cases}\begin{aligned}}
\newcommand{\casiend}{\end{aligned}\end{cases}}

\newcommand{\re}{\Re {\rm  e}}
\newcommand{\im}{\Im {\rm  m}}

\newcommand{\CiB}{{\rm CiB}_{p,\ell}^{(q_0,\xi)}}

\newcommand{\LG}{{\rm LG}}
\newcommand{\x}{{\bf x}}
\renewcommand{\eqref}[1]{Eq. (\ref{#1})}

\begin{document}
\title{The role of beam waist in Laguerre-Gauss expansion of vortex beams}

\author{Giuseppe Vallone}
\email{vallone@dei.unipd.it}
\affiliation{Department of Information Engineering, University of Padova, via Gradenigo 6/B, 35131 Padova, Italy}

\begin{abstract}
Laguerre-Gauss (LG) modes represent an orthonormal basis set of solutions of the paraxial wave equation.
LG are characterized by
two integer parameters $n$ and $\ell$ that are related to the radial and azimuthal profile of the beam. 
The physical dimension of the mode is instead determined by the beam waist parameter $w_0$:
only LG modes with the same $w_0$ satisfy the orthogonality relation.
Here, we derive the scalar product between two LG modes with different beam waists and show how this result can be exploited
to derive different expansions of a generic beam in terms of LG modes. 
In particular, we apply our results to the recently introduced Circular Beams, by deriving a
previously unknown expansion. We finally show how the waist parameter must be chosen in order to optimize
such expansion.
\end{abstract}

\maketitle

\section{Introduction} 
Laguerre-Gaussian (LG) beams
are exact solutions of the free-space paraxial wave equation 
in circular cylindrical coordinates. 
They form a complete base of orthogonal modes under which any paraxial optical field can be expanded.
Moreover, LG modes carry Orbital Angular Momentum (OAM) and their importance has raised together with the
recent developments~\cite{moli07nph} in applications of OAM for communication~\cite{vazi02prl,tamb12njp,bozi13sci,vall14prl}, 
imaging~\cite{furh05ope,mari12ope,hell15rmp} and fundamental physics~\cite{tamb11nph,damb13prx,damb14prl}.

LG beams are defined in term of two integer numbers $n$, $\ell\in \mathbb Z$ with $n\geq 0$:
the value of $n$ determines the radial profile of the LG mode, while 
$\ell$ is related to the OAM content of the beam~\cite{alle92pra}.
LG modes are also characterized by two dimensional quantities, the beam waist $w_0$
and its location $d_0$ along the propagation axis: these two quantities can be encoded in 
a single complex parameter $q_0\equiv -d_0+iz_0$, where $z_0=\kappa  w_0^2/2>0$ is the so called Rayleigh range and $\kappa$ is
the wavenumber~\cite{sieg86lasers}.
While $d_0$ can be arbitrarily changed by a translation on the propagation axis, 
the beam waist parameter $w_0$ determines the physical scale of the LG modes.
It is worth noticing that only LG with the same $q_0$ form a complete basis of orthogonal modes.

In the present work we will evaluate  the overlap integral of two LG modes with different complex parameters $q_0$ and $q'_0$. 
We will show that such overlap, besides the radial and angular parameters, 
depends only on a single adimensional complex variable given by a combination of $q_0$ and $q'_0$. 
We will show how such result can be exploited in finding new expansions
of generic beams in terms of LG modes and in particular we will apply our method to the recently introduced Circular 
Beams~\cite{band08opl,vall15opl}.

\begin{figure}
\centering\includegraphics[width=0.95\linewidth]{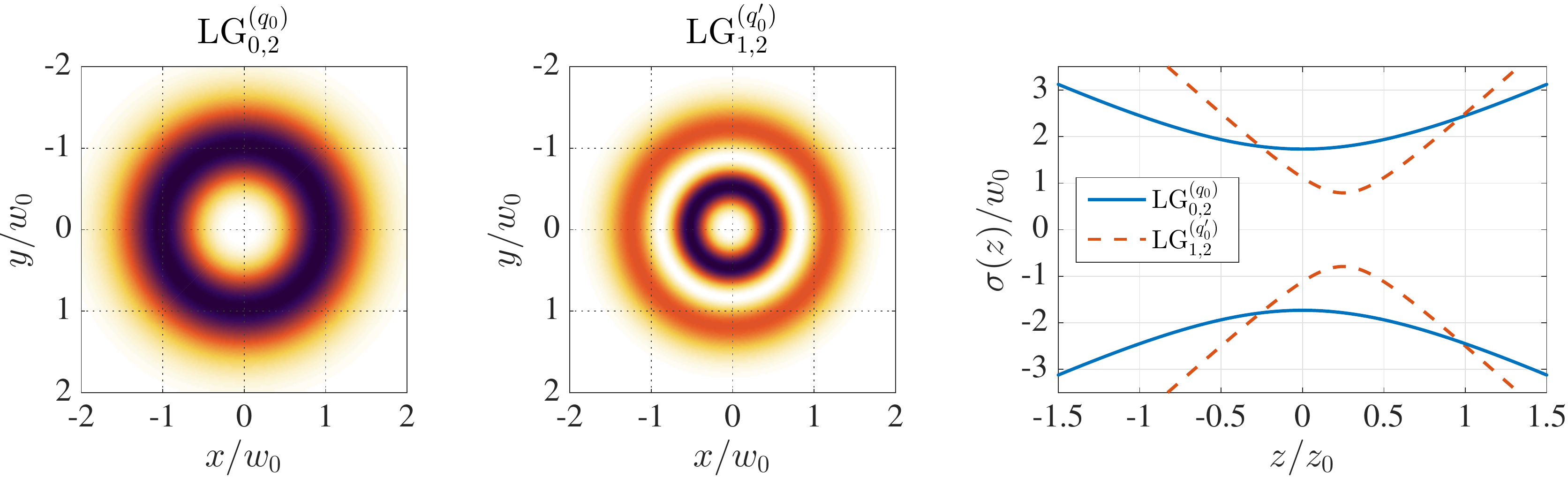}
\caption{Intensity at $z=0$ of two LG modes with different beam parameters $q_0$ and $q'_0$ and radial numbers $n=0$ and $n'=1$.
We used $q_0/z_0=i$ and $q'_0/z_0=i\gamma$ with $\gamma=(1+i)/4$. 
Despite that $n'\neq n$, the two beams are not orthogonal and their overlap integral 
 is $G^{(\gamma)}\simeq-0.46+0.23i$ (see \eqref{overlap}). In the right panel we also show the two beam radii $\sigma(z)$
as defined in \cite{vall16pra2}.}
\label{fig:LG}
\end{figure}
\section{Scalar product of LG modes} 
In this section we calculate the scalar product (defined by an overlap integral) between two LG modes with different beam parameters.
Our conventions for the normalized LG modes in cylindrical coordinates $\x=(r,\phi,z)$ are those used in Ref. \cite{vall15opl}.
The LG field is usually written in term of the 
Gouy phase $e^{i\zeta(z)}$
and the beam size $w(z)$. 
We here use an alternative notation, based on the  Siegman complex parameter $q(z)=z+q_0$.
By recalling that the Gouy phase and the beam size are  related to $q(z)$ by
$e^{i\zeta(z)}=i|q(z)|/q(z)$ and $w(z)=|q(z)|\sqrt{2/(\kappa z_0)}$,
the (normalized) LG mode can be explicitly expressed as
\beq 
\label{LG}
\begin{aligned}
\LG^{(q_0)}_{n,\ell}(\x)=&
\sqrt{\frac{ n!}{\pi(|\ell|+n)!}}
{(-\frac{q^*(z)}{q(z)})}^n
\frac{e^{-\frac{i\kappa r^2}{2q(z)}}}{r}
\times
\\
&
{(\frac{i \sqrt{\kappa z_0}}{q(z)}r)}^{|\ell|+1}
L_n^{|\ell|}\!\!\left(\frac{\kappa z_0r^2}{|q(z)|^2}\right)e^{i\ell\phi}\,,
\end{aligned}
\eeq
with $L_n^{|\ell|}(t)$  the generalized Laguerre polynomial.

We now derive the
scalar product $\braket{\LG^{(q'_0)}_{n',\ell'}}{\LG^{(q_0)}_{n,\ell}}$
between two  LG modes characterized by different beam parameters $q_0$ and $q'_0\equiv-d'_0+iz'_0$.
The scalar product between two generic beams $\Psi(\x)$ and $\Phi(\x)$ 
 is defined by the overlap integral  
$\braket{\Psi}{\Phi}=\int_0^{2\pi}\dd \phi\int_0^{+\infty}\dd r \,r\, \Psi^*(\x)\Phi(\x)$.
If $\Psi(\x)$ and $\Phi(\x)$ satisfy the
paraxial wave equation, $\braket{\Psi}{\Phi}$ 
is invariant under the following transformations applied to both beams:
the translation along the propagation axes $\mathcal T:\Psi(r,\phi,z)\rightarrow \Psi(r,\phi,z+d)$ and
the rescaling of the axis $\mathcal S:\Psi(r,\phi,z)\rightarrow \sqrt{s}\,\Psi(\sqrt{s}\,r,\phi,sz)$.
For the LG modes, the two above transformations can expressed in term of the complex parameter $q_0$ as 
 $\mathcal T:\LG^{(q_0)}_{n,\ell}\rightarrow \LG^{(q_0+d)}_{n,\ell}$
 and  $\mathcal S:\LG^{(q_0)}_{n,\ell}\rightarrow \sqrt{s}\,\LG^{(q_0/s)}_{n,\ell}$.
Then, the scalar product between two LG modes must depend only on a combination of $d_0$, $d'_0$, $z_0$ and $z'_0$
 that is invariant
under the two above transformations  applied to $q_0$ and $q'_0$.
A possible invariant combination is  the following (adimensional) complex variable:
\beq
\label{gamma}
\gamma =\frac{z'_0}{z_0}+i\frac{d'_0-d_0}{z_0}
\,.
\eeq
The  variable $\gamma$ can be equivalently defined by the implicit relation $q'_0=-d_0+i\gamma z_0$. 
A single constraint on $\gamma$, namely $\Re e (\gamma)>0$, 
is required due to the conditions $z'_0,z_0>0$.
It is easy to show that the parameter $\gamma$ defined by \eqref{gamma} is invariant 
under the two transformations $\mathcal T$ and $\mathcal S$.
Moreover, $\gamma$ has a simple physical interpretation: its real part is the square of the beam waist ratio and
it is thus related to the physical sizes of the two LG beams. Its imaginary part is the distance between the locations of the two beam waists measured
in units of $z_0$.

The LG modes defined in \eqref{LG} and with the same $q_0$ are orthonormal with respect to the above defined scalar product, namely
$\braket{\LG^{(q_0)}_{n',\ell'}}{\LG^{(q_0)}_{n,\ell}}=\delta_{n,n'}\delta_{\ell,\ell'}$.
On the other hand, as demonstrated in appendix \ref{Laguerre_integral},
the scalar product between two LG modes with different beam parameters $q_0$ and $q'_0$ 
is given by
\beq 
\label{overlap}
\braket{\LG^{(q'_0)}_{n',\ell'}}{\LG^{(q_0)}_{n,\ell}}
=\delta_{\ell,\ell'}G^{(\gamma)}_{n',n,\ell}\,,
\eeq
where
\begin{align}
\label{G}
\notag
G^{(\gamma)}_{n',n,\ell}
=&
\sqrt{\binom{|\ell|+n}{n}\binom{|\ell|+n'}{n'}}
{\tau_1}^{1+|\ell|}
\tau_2^{n}
\tau_3^{n'}
\times\\&
\ _2F_1(-n,-n',|\ell|+1;\frac{\tau_1^2}{\tau_2\tau_3})\,,
\end{align}
and the parameters $\tau_j$ depend on the complex variable $\gamma$ defined in \eqref{gamma} as follow:
\beq
\label{tau}
\tau_1\equiv\frac{2\sqrt{\Re e(\gamma)}}{1+\gamma^*}
\,,\quad
\tau_2\equiv
\frac{1-\gamma^*}{1+\gamma^*}
\,,\quad
\tau_3\equiv
\frac{\gamma-1}{1+\gamma^*}\,.
\eeq
We recall that the Hypergeometric polynomial $_2F_1$ in \eqref{G} is defined
by using the Pochhammer symbol
$(a)_k\equiv\Gamma(a+k)/\Gamma(a)$ as\footnote{We note that $(n)_k=\frac{(n+k-1)!}{(n-1)!}$ and 
$(-n)_k=(-1)^k\frac{n!}{(n-k)!}$ for $n,k\in\mathbb N$.}
$_2F_1(-n,b,c;z)=\sum^{n}_{k=0}\frac{(-n)_k(b)_k}{(c)_k}\frac{z^k}{k!}$.

If $q'_0\neq q_0$, LG modes with different radial numbers $n$ and $n'$
may become not-orthogonal (see Fig. \ref{fig:LG}), while modes with different OAM remains orthogonal due to the factor $\delta_{\ell,\ell'}$ in \eqref{overlap}. When $q'_0=q_0$, \eqref{overlap} reduces to the standard
orthogonality relation of the LG modes, as expected. Indeed, for $q'_0\rightarrow q_0$, we obtain $\gamma\rightarrow 1$ and 
$G^{(\gamma)}_{n',n,\ell}\rightarrow\delta_{n,n'}$. 

The scalar product between LG modes with different beam parameters
can be used to optimize the expansion of a generic beam. Indeed,
if a given expansion in term of  LG modes is known, e.g. $\Psi(\x)=\sum_{n,\ell}\psi_{n,\ell}\LG^{(q_0)}_{n,\ell}(\x)$,
by exploiting \eqref{overlap} the same field can be also expressed as
$\Psi(\x)=\sum_{n',\ell}\psi'_{n',\ell}
\LG^{(q'_0)}_{n',\ell}(\x)$
where
\beq
\label{newpsi}
\psi'_{n',\ell}=\sum_{n}\psi_{n,\ell}G^{(\gamma)}_{n',n,\ell}\,.
\eeq
While the expansion of $\Psi(\x)$ is unique at a given $q_0$
(indeed the LG modes represent a complete basis), by changing the beam parameter to $q'_0$,
a different expansion is found. It is worth noticing that, if the expansion is truncated such as
$\sum_{n\leq N}\sum_{|\ell|\leq L}\psi_{n,\ell}\LG^{(q_0)}_{n,\ell}(\x)$, the resulting beam may differently approximate the
original beam depending on the choice of the parameter $q_0$.
Then, the choice of the correct value of the beam waist size $w_0$ and its location $d_0$ for the
LG modes is essential for obtaining a faithful approximation in a truncated expansion.
We now apply the above considerations to the Circular Beams.

\section{Application to CiBs}
Circular Beams (CiBs) represent a very general solution of the paraxial wave equation
in cylindrical coordinate~\cite{band08opl}. They generalize many well known beam carrying OAM,
such as the elegant LG modes~\cite{wuns89josaA}, the Hypergeometeric-Gaussian (HyGG) beams~\cite{kari07opl, kari08ope}
or the optical vortex beam~\cite{berr04joa}.
The CiBs are determined by three complex parameters $\xi$, $q_0$ and $p$ and one 
integer parameter $\ell\in\mathbb Z$ related to the OAM content~\cite{vall16ope,vall16pra2}.
The parameter $\xi$ is related
to the beam shape and specific values identify some well known beams;
$q_0$ is related to the physical scale (similarly to the $q_0$ parameter of the LG modes) while
$p$ defines the radial index.

As demonstrated in~\cite{vall15opl}, the expansion of a CiB in terms of LG mode is written as:
\beq
\label{cib_exp}
\CiB(\x)
=\sum^{+\infty}_{n=0}C_n
\LG^{(q_0)}_{n,\ell}(\x)\,,
\eeq
with
\beq
\label{Cn}
C_{n}=C_0\frac{\xi^n(-p/2)_n}{\sqrt{n!(|\ell|+1)_n}}
\eeq
and
$C_0=(\!\ _2F_1[-\frac p2, -\frac{p^*}{2}, 1 + |\ell|,|\xi|^2])^{-1/2}$ is determined by normalization and
we used again the  Pochhammer symbol. 
In the notation of $C_n$, for simplicity we dropped the dependence on $\xi$, $p$ and $\ell$.
The above expansion is correct only for some subsets of the beam parameters that correspond to a square integrable CiB~\cite{vall15opl}. 
When $|\xi|<1$ the expansion in \eqref{cib_exp} is always well defined; if $|\xi|=1$ it is required that $\Re e(p)>-\ell-1$, while
when $|\xi|>1$ the parameter $p$ should be chosen as $p/2\in \mathbb N$.

It is worth noticing that in \eqref{cib_exp}, the CiB and the LG modes are defined with the same complex parameter $q_0$
(and the same $\ell$).
By using the result of \eqref{overlap}, we may now expand a CiB in terms of LG modes with
a different beam parameter $q'_0$.
As explained in the previous section, the new expansion can be written as
\beq
\label{new_exp}
\CiB(\x)=\sum^{+\infty}_{m=0}C'^{(\gamma)}_{m}{\LG^{(q'_0)}_{m,\ell}}(\x)\,,
\eeq
where $C'^{(\gamma)}_{m}=\sum_{n}C_{n}G^{(\gamma)}_{m,n,\ell}$ according to \eqref{newpsi}. 
As demonstrated in appendix \ref{new_expansion},
the new coefficients $C'_n$ can be explicitly evaluated as
\begin{align}
\label{C'}
\notag
\frac{C'^{(\gamma)}_{n}}{C'^{(\gamma)}_{0}}&= \tau_3^n
\sqrt{\binom{|\ell|+n}{n}}
\ _2F_1(-n,\frac {-p}2,|\ell|+1,\frac{\xi\tau_1^2}{\tau_3(\xi\tau_2-1)})\,,
\\
C'^{(\gamma)}_{0}&=C_{0}{\tau_1}^{1+|\ell|}(1-\xi\tau_2)^{p/2}\,.
\end{align}
with the $\tau_j$ parameters defined in \eqref{tau}.
\eqref{new_exp} and \eqref{C'} represent, to our knowledge, a previously unknown expansion of the Circular Beam in terms
of LG modes. 

\begin{figure}[t]
\includegraphics[width=0.95\linewidth]{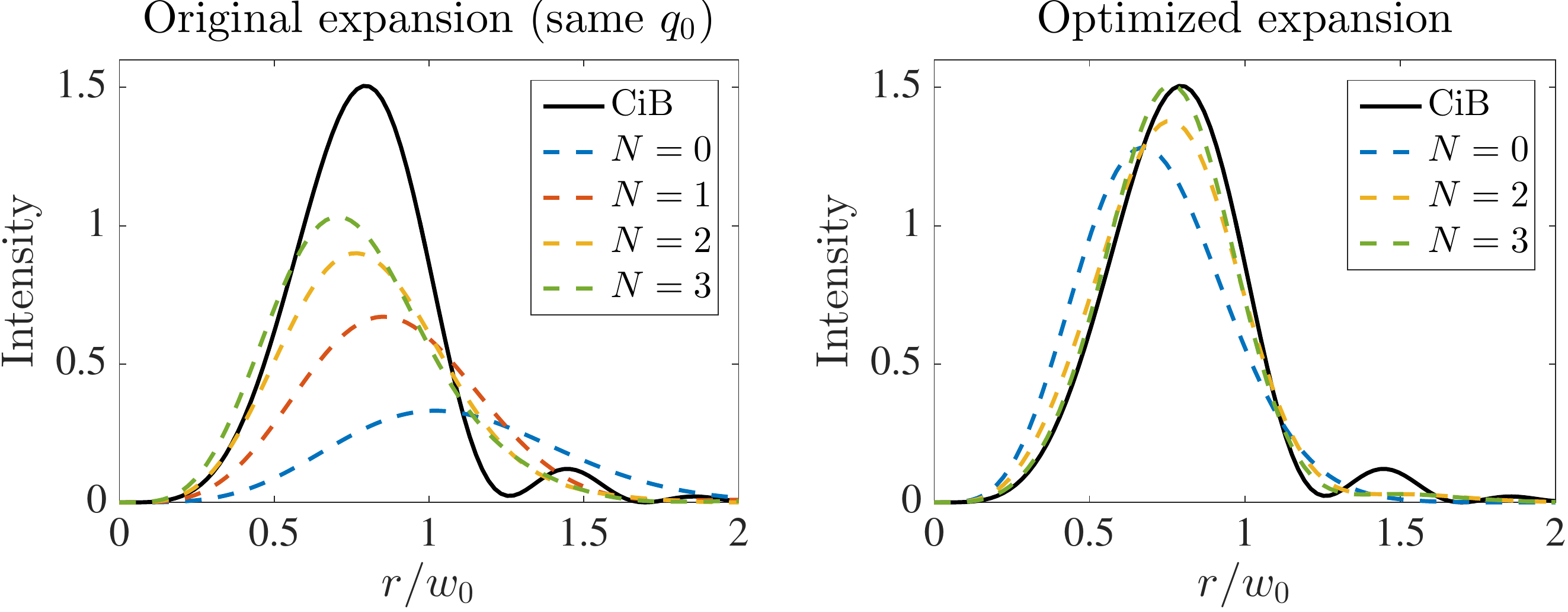}
\caption{Approximation by LG modes of a $\CiB$ with $\xi=1$, $\ell=-p=2$ and
$q_0/z_0=0.2+i$. The CiB is compared with the truncated expansions
$\sum^N_{n=0}C_n\LG^{(q_0)}_{n,\ell}$ and $\sum^N_{n=0}C'^{(\gamma_{\rm opt})}_n\LG^{(q'_0)}_{n,\ell}$
in the left and right graph respectively, for different value of $N$.
The optimized expansion is obtained by choosing $\gamma_{\rm opt}=1/3$ according to \eqref{gamm_simple}.
In the right figure the $N=0$ and $N=1$ expansions coincide since $C'^{(\gamma_{\rm opt})}_{1}=0$.}
\label{CibInLG}
\end{figure}

As already anticipated, by changing $\gamma$ (or equivalently $q'_0$), 
a truncated expansion of the form $T^{(\gamma)}_N=\sum^{N}_{k=0}C'^{(\gamma)}_{k}{\LG^{(q'_0)}_{n,\ell}}(\x)$ can be optimized. 
The larger is the value of the overlap
probability $P'_N(\gamma)=\sum^{N}_{k=0}|C'^{(\gamma)}_{k}|^2=|\braket{T^{(\gamma)}_N}{\CiB}|^2$, the better the 
truncated expansion approximates the original beam.
As an example, we will find  the value of 
$\gamma$ that optimize the expansion truncated to the first terms (namely for $N=0$):
in other words, we will look for  the value of $\gamma$ that maximize the
probability
\beq
\label{P'0}
P'_0(\gamma)=|C'^{(\gamma)}_0|^2={|C_{0}}|^2\cdot \left|\tau_1^{2+2|\ell|}(1-\xi\tau_2)^{p}\right|\,.
\eeq
Maximizing $P'_0$ corresponds in finding the LG mode with lowest radial number, namely $\LG^{(q'_0)}_{0,\ell}$,
that better approximates the CiB.
As shown in appendix \ref{optimization}, by defining the phase of $\xi$ as $2\varphi$, (i.e. $\xi=|\xi|e^{2i\varphi}$), the value of $\gamma$ that maximize $P'_0$ is given by
\beq
\label{gamma_opt}
\gamma_{\rm opt}=\frac{\gamma_+\cos\varphi-i \sin\varphi}{\cos\varphi -i\gamma_+\sin\varphi}\,,
\eeq
where
\beq
\label{gamma+}
\gamma_+=\frac{\alpha+\sqrt{\alpha^2+|\beta|^2(1-|\xi|^2)}}{\beta(1+|\xi|)}\,
\eeq
and
\beq
\label{alphabeta}
\begin{aligned}
\alpha=1+|\ell|+\Re e(p)\,,\qquad 
\beta=\frac{1+|\ell|}{|\xi|}-i\Im m(p)\,.
\end{aligned}
\eeq

A remarkable property is related to the value of the second expansion coefficient. 
When $\gamma=\gamma_{\rm opt}$, in the expansion of \eqref{new_exp} the coefficient of the LG mode with radial number $n=1$ is vanishing.
Indeed, By inserting $\gamma_{\rm opt}$ into
\eqref{C'} it is possible to demonstrate that
$C'^{(\gamma_{\rm opt})}_{1}=0$.

\begin{figure}[t]
\centering\includegraphics[width=0.7\linewidth]{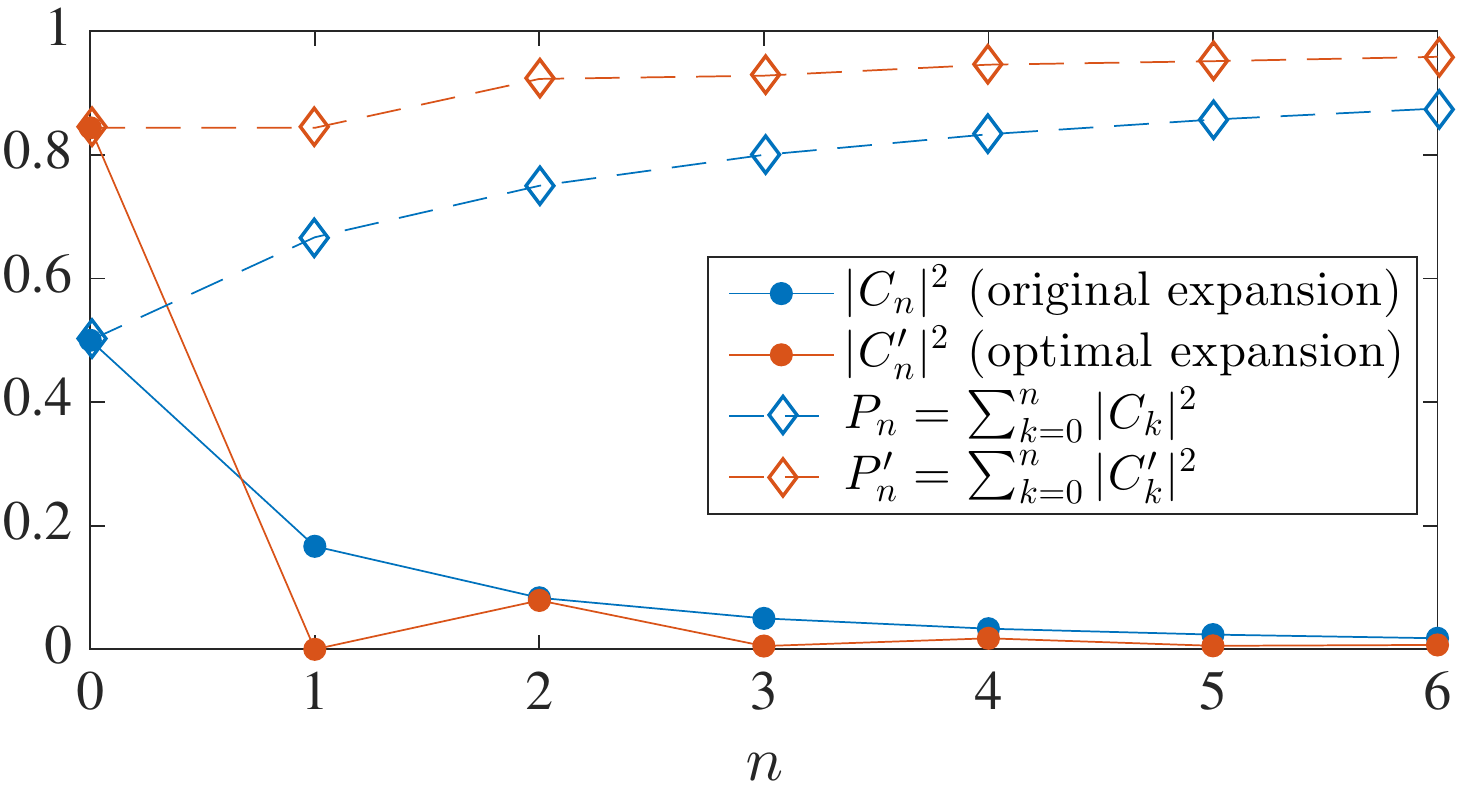}
\caption{ $|C_n|^2$ and overlap probability $P_n$ between the CiB and the truncated expansion (with the same parameters of Fig. \ref{CibInLG}).}
\label{Pn}
\end{figure}

\begin{figure}[t]
\centering\includegraphics[width=0.85\linewidth]{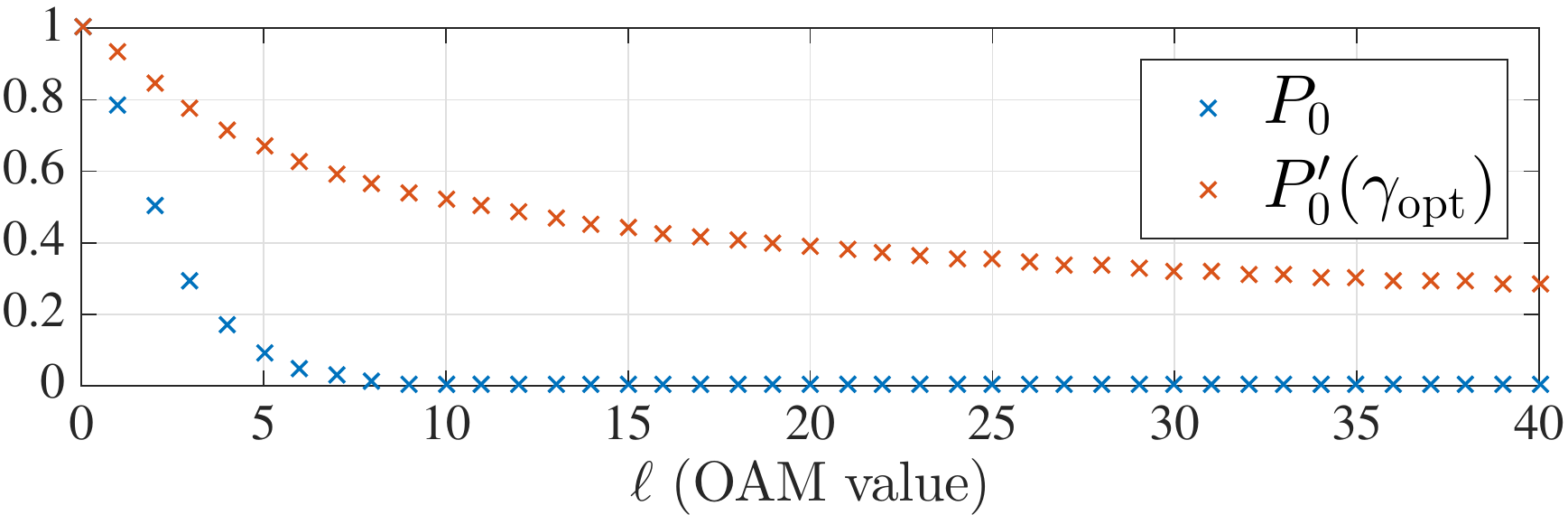}
\caption{$P_0$ and $P'_0(\gamma_{\rm opt})$ for the $\xi=1$, $p=-|\ell|$ beams.}
\label{P0}
\end{figure}

An important sub-case is obtained for $|\xi|=1$, namely when the CiBs reduce to the
(generalized) HyGG modes~\cite{kari07opl,kari08ope,vall15opl}. 
As demonstrated in~\cite{kari07opl,vall15opl}, the
 HyGG modes have a very simple profile at the plane $z=d_0+iz_0\frac{\xi-1}{\xi+1}$, given by
 ${\rm HyGG}\propto r ^{p+|\ell|}\exp(-\frac{\xi+1}{2\xi}\frac{r^{2}}{w_0^2}+i\ell\phi)$ and thus can be easily generated experimentally.
For such beams, the optimal $\gamma_{\rm opt}$ reduces to a very  simple expression
\beq
\label{gamm_simple}
\gamma_{\rm opt}=1+\frac{2p\xi}{{2+2|\ell|+p^*-p\xi}}\,,
\qquad \text{for }|\xi|=1\,.
\eeq
As an example, in Fig. \ref{CibInLG} and \ref{Pn} we show the truncated approximation of a $\CiB$ with $\xi=1$, $\ell=-p=2$ and
$q_0/z_0=0.2+i$ for two values of $\gamma$. In the left plot of Fig. \ref{CibInLG} we used  LG modes with the same $q_0$
of the CiB (corresponding to $\gamma=1$).
In the right plot of Fig. \ref{CibInLG} we used $\gamma=1/3$ (i.e. $q'_0/z_0=0.2+i/3$), representing 
the optimal value obtained by \eqref{gamm_simple} and corresponding to LG modes with
a beam waist shrank by a factor $\sqrt{3}$ with respect to the $\gamma=1$ case. 
The figures clearly show that with the optimized value of $\gamma$, the truncated expansion better approximates the
CiB for lower values of $N$. Indeed, in this specific case, more than $84\%$ of the CiB energy is contained in the first LG mode 
with optimized $q'_0$ ($\LG^{(q'_0)}_{0,2}$), while to obtain the same energy it necessary to sum the first six LG modes with the original
$q_0$ parameter. This improvement holds also for higher values of $\ell$ as shown  in Fig. \ref{P0}.

\section{Conclusions}
The scalar product between two LG modes with different beam parameters $q_0$ and $q'_0$
was explicitly evaluated (see \eqref{overlap})  and
it was used to find new expansions of generic beams in terms of LG mode. 
By the above results, a previously unknown expansion of the Circular Beams is obtained (see \eqref{new_exp} and \eqref{C'}).
Finally, the value of the LG beam parameter $q'_0$ that optimizes the overlap probability of the  $\LG_{0,\ell}$ mode
with the CiB has been found (see \eqref{gamma_opt}).
Our results have important applications in OAM generation, manipulation and detection, since they allow to
precisely determine and optimize the expansion of a generic beam in terms of LG modes, the fundamental beam carrying OAM.
The case studied in \eqref{gamm_simple} is particularly relevant for experiments
since CiBs with $|\xi|=1$ can be easily  generated experimentally \cite{kari07opl,vall16ope}.

\acknowledgements
We thanks Paolo Villoresi of the University of Padova, Filippo Cardano and Lorenzo Marrucci of the University Federico II of Napoli
for useful suggestions and discussions. Our work was supported by the ``Progetto di Ateneo PRAT 2013, OAM in free space: a new resource for QKD (CPDA138592)'' of the University of Padova.

\appendix
\section{Laguerre polynomial integral}
\label{Laguerre_integral}
We now demonstrate the result stated in \eqref{overlap}. Integration over the angular variable give rise to the $\delta_{\ell,\ell'}$ factor
in \eqref{overlap}.
After changing variables in the radial integration,
 $\braket{\LG^{(q'_0)}_{n',\ell'}}{\LG^{(q_0)}_{n,\ell}}$ is proportional to a single integral 
involving two Lauguerre polynomials as
$\int^{\infty}_0\dd t\,\, t^{\ell}
e^{-c t}
L_{n}^{|\ell|}(a t)
L_{n'}^{|\ell|}(b t)$
with $c=\frac{i}{2}\frac{q'^*_0-q_0}{q_0 q'^*_0}$, 
$a=\frac{z_0}{|q_0|^2}$ and $b=\frac{z'_0}{|q'_0|^2}$. We note that $\re(c)=\frac{a+b}{2}>0$.
\eqref{G} is derived by evaluating the above integral as:
\beq
\label{LGintegral}
\begin{split}
&\frac{c^{\ell+1}}{\ell!}\int^{+\infty}_0\dd t\,\, t^{\ell}
e^{-c t}
L_{n}^{\ell}(a t)
L_{n'}^{\ell}(b t)=\binom{\ell+n}{\ell}\binom{\ell+n'}{\ell}\times
\\
&(1-\frac{a}{c})^{n}(1-\frac{b}{c})^{n'}
\ _2F_1(-n,-n',\ell+1;\frac{ab}{(c-a)(c-b)})\,.
\end{split}
\eeq
\eqref{LGintegral}, holding when $\Re e(c)>0$ and  $n$, $n'$ and $\ell$ are
non negative integers,
is obtained by expressing the two Laguerre polynomials in term of their generating function
as 
\beq
L^{\ell}_n(x)=\left.\frac{1}{n!}(\frac{\dd}{\dd \eta})^{n}\left[\frac{1}{(1-\eta)^{\ell+1}}\exp(-\frac{x}{1-x}\eta)\right]\right|_{\eta=0}
\eeq
and then evaluating the Gaussian integral.
The final result is demonstrated by recalling that 
$w(\eta,x)={(1-\eta)^{b-c}}/{(1-\eta+x\eta)^b}$
is the generating function of the hypergeometric polynomial, namely 
\beq\left.\frac{d^nw(\eta,x)}{\dd \eta^n}\right|_{\eta=0}=\frac{(c+n-1)!}{(c-1)!}\ _2F_1(-n,b,c;x)\,.
\eeq

\section{New expansion of the CiB}
\label{new_expansion}
According to \eqref{newpsi}, the coefficient of the new expansion are given by $C'^{(\gamma)}_{m}=\sum_{n}C_{n}G^{(\gamma)}_{m,n,\ell}$.
By using the coefficient $C_n$ defined in \eqref{cib_exp} we obtain:
\begin{align}
\notag C'^{(\gamma)}_{n'}=&
C_{0}\sum_{n}
\frac{\xi^n(-p/2)_n}{\sqrt{n!(|\ell|+1)_n}}
G^{(\gamma)}_{n',n,\ell}
\label{C'_appendix}
\\
=&C_{0}
{\tau_1}^{1+|\ell|}\tau_3^{n'}\sqrt{\binom{|\ell|+n'}{n'}}
\times\\&
\sum^{\infty}_{k=0}\frac{(-n')_k}{(|\ell|+1)_k k!}(\frac{-\tau_1^2}{\tau_2\tau_3})^k
\sum_{n=k}^{+\infty}
\frac{(\xi\tau_2)^n}{(n-k)!}(\frac{-p}{2})_n\,.
\notag
\end{align}
In the last equality we 
used the definition of the Hypergeometric polynomial 
and we switched the sums over $k$ and $n$. We now exploit the following relation
$\sum^{+\infty}_{n=k}
\frac{z^n}{(n-k)!}
(-a)_n
=z^k(1-z)^{a-k}
(-a)_k
$,
holding when $|z|<1$ and $\forall a$ or when $|z|>1$ and $a\in \mathbb N$.
When $|\xi|\leq 1$ we have $|\tau_2|<1$ (since $\re(\gamma)>0$) and  the  sum over $n$ in the r.h.s. of \eqref{C'_appendix} always converges
to $(1-\xi\tau_2)^{p/2}(-p/2)_k(\frac{\xi\tau_2}{1-\xi\tau_2})^k$.
When $\xi> 1$, square integrability of the CiB requires $p=2m$ with $m\in\mathbb N$:
in this case the sum over $n$ in \eqref{C'_appendix} converges to the same value
since all the terms with $n>m$ vanish
due to the factor $(-p/2)_n=(-m)_n$. Then we obtain
\beq
\label{C'_appendix2}
\begin{aligned}
C'^{(\gamma)}_{n}
=&
C_{0}
{\tau_1}^{1+|\ell|}(1-\xi\tau_2)^{p/2}\tau_3^{n}\sqrt{\binom{|\ell|+n}{n}}
\times\\&
\quad\sum^{n}_{k=0}\frac{(-n)_k(-p/2)_k}{(|\ell|+1)_k}\frac{1}{k!}(\frac{\xi\tau_1^2}{\tau_3(\xi\tau_2-1)})^k\,.
\end{aligned}
\eeq
The sum over $k$, due to the term $(-n)_k$, is limited to $n$ 
and it converges to the hypergeometric polynomial. We have thus demonstrated \eqref{C'} of the main text.

\section{$\gamma$ optimization}
\label{optimization}
The value of $\gamma$ that maximize $P'_0(\gamma)$ of \eqref{P'0} is found by solving the equation
$|\gamma|^2(\xi+1)+2i\im(\gamma)-2\xi(1+\frac{p}{|\ell|+1})\re(\gamma)+\xi-1=0$
obtained by imposing $\frac{\partial P'_0}{\partial \gamma}=0$. When $\xi=|\xi|$, the
above equation is  solved by
$\gamma_\pm=\frac{\alpha\pm\sqrt{\alpha^2+|\beta|^2(1-|\xi|^2)}}{\beta(1+|\xi|)}$,
with $\alpha$ and $\beta$ defined in \eqref{alphabeta}.
When $|\xi|\leq 1$ only the $\gamma_+$ solution is physical since $\gamma_-$ does not satisfy the condition $\Re e(\gamma)>0$.
For $|\xi|>1$ the  solution $\gamma_-$ is physical but it becomes a saddle point and not a maximum for $P_0$. Then, when $\xi=|\xi|$, the
value of $\gamma$ that maximize $P_0$ is given by $\gamma_+$ defined above.

When $\xi\neq |\xi|$ the solution of the maximizing equation can be found by exploiting 
the properties of the CiBs under generic optical transformation defined by the ABCD law~\cite{Saleh-Teich}.
As detailed in~\cite{vall16ope}, the beam parameters $\xi$ and $q_0$ are 
transformed as $\xi\rightarrow\xi_1=\xi\frac{C q^*_0+D}{C q_0+D}$ 
and
$q_0\rightarrow q_1=\frac{A q_0+B}{C q_0+D}$
when the CiB passes through an optical system defined by the matrix
$M=(\begin{smallmatrix} A & B
\\
C & D
\end{smallmatrix})$. If $\xi\neq|\xi|$ we may use the ABCD law to transform $\xi$ into $\xi_1$ such that $\xi_1$ is real and positive.
If $\xi=|\xi|e^{2i\varphi}$, the required matrix $M_\xi$ has parameters $A=D=1$, $B=0$ and $C=\frac{1}{d_0+z_0\cot\varphi}$:
for such values we have $\xi_1=|\xi|$. In this system the LG expansion is optimized with $q'_1=\re(q_1)+i\gamma_+\im(q_1)$.
By inverting the transformation with $M^{-1}_\xi$, the optimal $q'_0$ can be found, from which
$\gamma_{\rm opt}=-i\frac{q'_0+d_0}{z_0}=\frac{\gamma_+\cos\varphi-i \sin\varphi}{\cos\varphi -i\gamma_+\sin\varphi}$.
We have thus demonstrated \eqref{gamma_opt} of the main text.
%

\end{document}